\documentclass[12pt]{article} \textheight=23cm \textwidth=16cm
\usepackage{amsmath} 
\usepackage{amsfonts}
\usepackage{amssymb} 
\usepackage{graphicx}

\def\beq{\begin{equation}}
\def\eeq{\end{equation}}
\def\eea{\end{eqnarray}}
\def\bea{\begin{eqnarray}}

\def\hsm{h_{\rm SM}}
\def\mhsm{m_{h_{\rm SM}}}
\def\tev{\, {\rm TeV}}
\def\gev{\, {\rm GeV}}
\def\mev{\, {\rm MeV}}

\def\xfb{\, {\rm fb}}
\newcommand{\gsim}{\lower.7ex\hbox{$\;\stackrel{\textstyle>}{\sim}\;$}}
\newcommand{\lsim}{\lower.7ex\hbox{$\;\stackrel{\textstyle<}{\sim}\;$}}

\def\slashchar#1{\setbox0=\hbox{$#1$}           
   \dimen0=\wd0                                 
   \setbox1=\hbox{/} \dimen1=\wd1               
   \ifdim\dimen0>\dimen1                        
      \rlap{\hbox to \dimen0{\hfil/\hfil}}      
      #1                                        
   \else                                        
      \rlap{\hbox to \dimen1{\hfil$#1$\hfil}}   
      /                                         
   \fi}                                         %

\begin{document}


\pagestyle{empty}
\parskip=0.13cm

\catcode`\@=11
\def\lesssim{\mathrel{\mathpalette\vereq<}}
\def\gtrsim{\mathrel{\mathpalette\vereq>}}
\def\vereq#1#2{\lower3pt\vbox{\baselineskip0pt \lineskip0pt
\ialign{$\m@th#1\hfill##\hfill$\crcr#2\crcr\sim\crcr}}}
\catcode`\@=12

\def\alt{\lesssim}
\def\agt{\gtrsim}
\bigskip

\begin{center}
{\Large \bf Extra Dimensions and the Universal Suppression \\ 
of Higgs Boson Observables at High Energy Colliders}

\bigskip

James D. Wells

Physics Department

University of California

Davis, CA 95616

\end{center}

\begin{abstract}

Precision electroweak data suggests the existence of a light Standard
Model Higgs boson.  Its width is very narrow by an accident of the
fermion mass hierarchies, and so slight perturbations to the theory
induce dramatic changes in the phenomenology of direct Higgs boson
production.  Several theory ideas motivated by low-scale extra
dimensions lead to a universal rate suppression of Higgs boson
observables.  The Tevatron and LHC will have difficulty accruing
compelling evidence for small universal suppression of rate
observables, and even more difficulty discerning the underlying cause.
A high energy $e^+e^-$ collider would have relatively little trouble.

\end{abstract}

\section*{The narrow Higgs boson}

The most effective searches to date 
for the Higgs boson in the Standard Model (SM)
come from $e^+e^-\to \hsm Z$ experiments at LEPII. They have put
a 95\% C.L.\ numerical limit on the Higgs boson mass of
$m_{\hsm}>114.1\gev$~\cite{unknown:2001xw}.  
Meanwhile, precision electroweak analyses
based mostly on LEP, SLD and Tevatron data have put indirect limits on the
SM Higgs boson mass~\cite{Abbaneo:2001ix}:
\bea
\label{log equation}
\log_{10}(\mhsm/{\rm GeV})=1.94^{+0.21}_{-0.22}.
\eea
The 95\% C.L.\ upper bound is $222\gev$~\cite{Abbaneo:2001ix}.  
The closer the Higgs boson is to the $114.1\gev$ lower bound the
better it can account for the precision electroweak data.

If the Higgs boson is less than about $2m_W$ its width is less
than $100\mev$.  Such a narrow Higgs width is due to the SM accident of
all fermions having mass less than $5\gev$ except the top quark, which
is too heavy for the light Higgs to decay into anyway. The narrowness
of the Higgs width makes the Higgs phenomenology
highly susceptible to subtle interactions the Higgs  might have with
other exotic states.  

\section*{Extra dimensions and the Higgs boson}

Large extra dimensions felt by gravity can generate
the otherwise unnaturally large 
hierarchy between the Planck scale and the weak 
scale~\cite{Arkani-Hamed:1998rs}.  This would
effectively preclude the existence of any fundamental scale far above the
weak scale.  The implications of this idea to the Higgs sector
are significant.  

For one, supersymmetry
might not be needed to solve the quadratic sensitivity problem, since there
would be no scales high enough to cause us concern. Superpartners would
not be around, and perhaps more important to this discussion, there would be 
no obvious need for a second Higgs doublet as there is in supersymmetry.

Second, the existence of a large higher-dimensional
bulk space
can be important in ways that were never relevant in high-scale 
extra dimensional theories. For example, the mixing of the Higgs boson
with graviscalars can radically change Higgs 
phenomenology~\cite{Giudice:2000av}. Likewise,
the decay of the Higgs boson into bulk states (states that feel the large
dimensions with gravity) can change the phenomenology.

Finally, when there is only a few TeV of room to solve all possible
problems in particle physics there are sure to be states living just
above the weak scale that can mix with the Higgs boson. For example,
to solve the proton decay problem may require the introduction of a new
protecting gauge symmetry, such as baryon number or lepton number.
Although it is not necessary to introduce spontaneous symmetry
breaking in the conventional sense for all of these new 
symmetries, it is likely that at least one SM singlet
scalar exists for such purposes which would then
mix with the Higgs doublet of the SM and change its phenomenology.

\section*{Universal suppression via invisible width}

There are many theoretical examples demonstrating the possibility
that the Higgs boson
can decay into states that are not detectable by our current technology
(for review, see e.g., \cite{Martin:1999qf}).
In the extra 
dimensional framework there are two interesting and unrelated ideas that 
could lead to an invisible width of the Higgs boson while
not affecting Higgs boson interactions with any other state.

The first idea comes from neutrinos in the bulk~\cite{Arkani-Hamed:1998vp}.  
The small neutrino masses
that we now believe must exist could be entirely due to Dirac
masses that are suppressed by a large volume factor.  Since the right-handed
neutrino is not charged under any SM gauge symmetry, it is
the most likely of all SM fields to join gravity in the bulk. When
performing the KK reduction of this state, one finds naturally 
that the neutrino Dirac
mass can be in the sub-eV range from bulk volume suppression.  

In the 4-dimensional picture, the
sum over the enormous number of KK states accessible in 
$h\to \nu_L\bar\nu^{(i)}_R$ decays contributes substantially
to the Higgs width:
\bea
\Gamma_{\rm inv}(h\to \nu_L\bar\nu_R^{(i)}) \simeq  (1\mev)\, 20^{6-\delta}
\left(\frac{m_h}{150\gev}\right)^{1+\delta} 
\left(\frac{m_\nu^2}{10^{-3}\, {\rm eV}^2}\right)
\left(\frac{3\tev}{M_D}\right)^{2+\delta},
\eea
where $\delta$ is the number of extra dimensions and $M_D$ is the fundamental
scale of gravity.
Each KK state is long-lived, so the above expression leads to a potentially
significant invisible width of the Higgs boson~\cite{Arkani-Hamed:1998vp}.

There is another source of invisible width coming from 
kinetic mixing of the Higgs boson with graviscalars in the low
scale gravity action~\cite{Giudice:2000av,Antoniadis:2002ut},
\bea
S=-\xi\int d^4x \sqrt{-g_{\rm ind}}R(g_{\rm ind})H^\dagger H.
\eea
Expanding this action out in terms of the physical graviton and
graviscalar fields, substituting $H=e^{i\eta}(v+h)/\sqrt{2}$, and 
utilizing the equations of motion, we find a mixing term between
the Higgs and graviscalar fields $\varphi^{(\vec n)}_G$ 
induced in the lagrangian~\cite{Giudice:2000av},
\bea
{\cal L}_{\rm mix}=-\frac{2\xi vm^2_h}{M_P}\sqrt{\frac{3(\delta -1)}{\delta+2}}
\, h\sum_{\vec n} \varphi^{(\vec n)}_G.
\eea

The mass gap spacing is sufficiently small between the KK excitations
of the graviscalars that we can talk meaningfully about oscillations of
the Higgs field into graviscalars. This oscillation is equivalent to giving 
the Higgs an invisible width,
\bea
\Gamma_{\rm inv}(h\to \varphi^{(\vec n)}_G)\simeq (8\mev)\, 20^{2-\delta}\, 
\xi^2\, S_{\delta -1}\frac{3(\delta -1)}{\delta+2} 
\left(\frac{m_h}{150\gev}\right)^{1+\delta}\left(
\frac{3\tev}{M_D}\right)^{2+\delta},
\eea
where $S_{\delta -1}$ is the surface area of a $\delta$-dimensional
unit radius sphere.

Having established that extra dimensional theories have several
ways in which they could produce an invisible decay width to the
Higgs boson, we now analyze the effect it has on Higgs boson observables
at high energy colliders. 

Each rate observable is a multiplication
of a particular Higgs boson production cross-section $\sigma_i(h)$
($i$ labels the other final state particles in the process)
times the relevant branching fraction $B_j(h)$ of the final
state of the Higgs we are interested in.  Ignoring the complications
of experimental efficiencies and background reduction techniques, 
what will be measured is the total rate
of a particular signal,
\bea
R_{ij}(h)=\sigma_i(h)B_j(h).
\eea
Since the width of the light Higgs boson is smaller
than the experimental resolution,
the only observables possible at the Tevatron and LHC
are these rate observables $R_{ij}(h)$ and 
simple mass reconstruction of the Higgs boson. 

An invisible width will not change the production cross section of the Higgs.
However, the branching ratios into observable SM states do change:
\bea
B_i=\frac{\Gamma_i}{\Gamma_{\rm tot}+\Gamma_{\rm inv}} 
= B_i^{\rm SM}(1-B_{\rm inv}).
\eea
It is important to emphasize that the reduction in all SM branching
fractions is independent of the final state, and is given by the universal
value $1-B_{\rm inv}$.

We therefore conclude that all SM rate observables
involving the Higgs boson will be reduced by the same universal
constant
\bea
\label{inv rate}
R_{ij}(h)=R^{\rm SM}_{ij}(h)(1-B_{\rm inv}).
\eea
Using only pure rate observables, it is therefore impossible to determine
whether the cross-section or branching fraction (or both) is the cause
of the reduction in rate.

\section*{Universal suppression via singlet mixing}

Suppose the low-energy effective theory contains
another real singlet field $S$, which does not couple to
any SM state, but mixes with the real SM Higgs field according to the
lagrangian,
\bea
{\cal L}  =  \frac{1}{2}(\partial_\mu \hsm)^2-\frac{1}{2}\mhsm^2\hsm^2 
+\frac{1}{2}(\partial_\mu S)^2-\frac{1}{2}m_{S}^2
S^2 - \mu^2\hsm S + {\cal L}_I^{\rm SM}(\hsm) +\ldots
\eea
where ${\cal L}_I^{\rm SM}(\hsm)$ contains the well-known interactions
between the SM Higgs and other SM particles.

One possible origin of the field $S$ could be from symmetry breaking of
a TeV scale $U(1)$ gauge symmetry.  Extra U(1)'s are ubiquitous at the
string scale in string model building~ and discrete remnants
are very useful in protecting
symmetries to all orders such as may be required to solve the proton
decay issue in low-scale extra dimensional theories.  Although all or some
of the
U(1) symmetries might possibly
be broken by other means, condensing scalar fields
obviously remain a viable and attractive means to accomplish symmetry breaking.

The coupling of a charged, complex singlet
field $\Phi_S=S+iA_S$ to the SM states would be
\bea
{\cal L}(\Phi_S) = \sum \frac{|\Phi_S|^2}{\Lambda^2}{\cal O}_i^{\rm SM} 
 + m_\Phi^2|\Phi_S|^2 
-\lambda_1|\Phi_S|^4-\lambda_2 |\Phi_S|^2|H_{\rm SM}|^2
+ \cdots
\eea
where ${\cal O}_i^{\rm SM}$ are all the marginal operators of the SM.
After spontaneous symmetry breaking the $A_S$ field is eaten by the extra
gauge symmetry and the $S$ is left over to mix with the remaining
real degree of freedom $h_{\rm SM}$ from the SM Higgs doublet $H_{\rm SM}$. 
In the limit $\Lambda\to\,{\rm large}$ the coupling to SM fermions and
gauge fields goes to zero, recovering the limit of no interactions
of $S$ with SM fields except through its mixing with the SM Higgs.

The mass eigenstates $\sigma$ and $h$ are obtained by 
replacing $S$ and $\hsm$ with
\bea
S  & \to & \cos\omega\, \sigma - \sin\omega\, h \nonumber \\
\hsm & \to & \sin\omega\, \sigma +\cos\omega\, h \nonumber 
\eea
where
\bea
\tan 2\omega & = &\frac{2\mu^2}{m_{S}^2-\mhsm^2},~~~~{\rm and} \nonumber\\
m^2_{\sigma,h} & = &\frac{1}{2}\left[ m_{S}^2+\mhsm^2\pm
\sqrt{(m_{S}^2-\mhsm^2)^2+4\mu^4}\right] . \nonumber
\eea\nonumber \\

After diagonalizing the Lagrangian the new Higgs boson $h$ couples
like the SM Higgs boson except its couplings to SM fields 
are reduced by a factor
of $\cos\omega$.  Since $S$ by construction has zero or, more
realistically, negligible interactions with the SM states,
the component of the mass
eigenstate $h$ that overlaps with $S$ does
not contribute to $h$ phenomenology.
The field $\sigma$ picks up interactions by virtue
of its overlap with $\hsm$ of strength $\sin\omega$.  Therefore,
$\sigma$ also has an effect on the precision electroweak analysis.

\begin{figure}[bt]
\begin{center}
\includegraphics*[totalheight=2.5in]{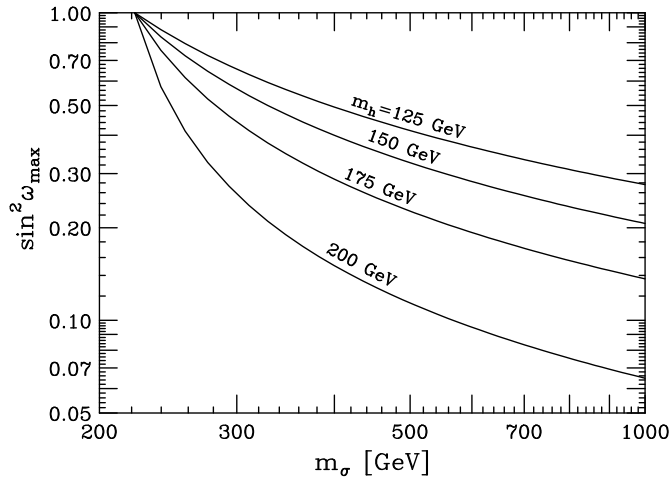}
\caption{95\% C.L. limits on heavier singlet state and
its mixing with the Higgs boson from precision electroweak
data fits.  For example, if the Higgs boson is found at $175\gev$,
the value of $\sin^2\omega$ must be less than about 0.14 for 
$m_\sigma\simeq 1\tev$.}
\label{s2omega}
\end{center}
\end{figure}

We can quickly see the interesting effects of a heavy $\sigma$
by the following rough analysis.  If there is non-zero $\sin^2\omega$, the mixed singlet will also contribute
at one-loop to the electroweak precision observables.  
The $\log \mhsm$ in 
Eq.~\ref{log equation} now becomes
\bea
\log (\mhsm/{\rm GeV}) 
\longrightarrow \log (m_h/{\rm GeV}) +\sin^2\omega \log m_\sigma/m_h. 
\eea
For $m_\sigma > m_h$ the upper limit on the Higgs mass is never greater
than what it is in the SM. 
If the singlet
mass is too high, the total $\chi^2$ 
fit would be unacceptably high. 
In Fig.~\ref{s2omega} I estimate the 95\% C.L limit of the maximum
value of $\sin^2\omega$ allowed as a function of $\sigma$ mass for
various Higgs masses.  This plot gives a sense of how large
the mixing is allowed to be for very heavy singlets inaccessible to 
high energy collider production. In that case only the lighter $h$
field will be accessible at the colliders, and all information must
come from its study.

The $h$ field will decay with exactly the same branching fractions as the 
SM Higgs, however its production cross section will
be suppressed by a factor of $\cos^2\omega$.  For rate observables
this translates into a suppression compared to the expected
SM rates,
\bea
\label{omega rate}
R_{ij}(h)=R^{\rm SM}_{ij}(h)(1-\sin^2\omega).
\eea
Notice that again the suppression factor is a universal,
process independent suppression factor.  By measuring any number of
SM rate observables it is impossible to distinguish $B_{\rm inv}$
from $\sin^2\omega$ (cf.\ eq.~\ref{inv rate} with eq.~\ref{omega rate}).

%

\section*{Challenges for present and future colliders}

The power of experiment derives from its ability 
to collect otherwise inaccessible experience
about nature, thus enabling us to refine our theories of natural law.
An important measure of progress is the continuing ability
to eliminate competing ideas.  

In the above paragraphs I have outlined several generic ways in which
a light Higgs boson can have the same signal rates
as the SM Higgs except with a universal rescaling factor.
That discussion highlights three main challenges that all
present and future colliders should seriously consider:
\begin{itemize}
\item To what sensitivity can experiment measure deviations of the $R_{ij}(h)$
rate observables compared to the SM expectation?
\item How much evidence can one accrue that all $R_{ij}(h)$
rate observables are rescaled by a universal suppression factor?
\item Is it possible to distinguish a universal suppression factor
in the branching fraction from a universal suppression
factor in the production cross-section?
\end{itemize}

At the Tevatron, obtaining a $3\sigma$ evidence for a SM Higgs boson
with mass up to $180\gev$ will
require summing over all possible rate observables in
$30\xfb^{-1}$ per experiment
of integrated luminosity~\cite{Carena:2000yx}.  
Significance of discovery and sensitivity
to deviations in any one channel will be a significant challenge.
One could attempt to measure a non-SM rate observable associated with the
invisible decay of the Higgs boson, say, in 
$p\bar p\to Zh\to l^+l^-+\slashchar{E_T}$.  However, even if the invisible
branching ratio were 100\% it is unlikely that Tevatron would find
a significant signal for this very challenging rate 
observable~\cite{Martin:1999qf}.  
It only gets worse when the
branching fraction decreases.

The LHC may be able to  measure SM rate observables
in many different channels.  The approachable channels
depend on the mass of the Higgs boson.  For $m_h=130\gev$
it may be possible to measure the rate observables in 
$gg\to h\to \gamma\gamma$, $gg\to h\to WW\to l^+{l'}^-+\slashchar{E_T}$,
$q\bar q\to Wh\to l b\bar b+\slashchar{E_T}$, $VV\to h\to WW^*$ and
$q\bar q\to Wh\to WWW\to 3l+\slashchar{E_T}$. The measurement
of all rate observables at the LHC will never yield a result
with better than 10\% uncertainty with the possible exception
of $VV\to h\to WW^*$ with $200\xfb^{-1}$ of data~\cite{Rainwater:2002hm} .
Furthermore,
the QCD uncertainties suggest significant challenges in the
theoretical prediction of rates at hadron colliders~\cite{LHC Note}.

It has been shown~\cite{Zeppenfeld:2000td}
that if one makes several assumptions about how
the Higgs boson interacts with SM
particles and makes the assumption that there is very little
interaction with anything else (not satisfied in large
invisible width case), one
might be able to infer the total width of the SM Higgs to within
10-20\% for $120\gev<m_h<250\gev$ with $100\xfb^{-1}$ of data.  
As the production rate goes down, indicating
a larger invisible width or more mixing with a singlet, the
statistical and systematic errors in inferring the Higgs width will
increase.  

Even in the ideal case that
a large deviation is found for each rate
observable, and the evidence points to a universal suppression,
it still looks difficult to distinguish between the radically different 
prospects of a Higgs with an invisible decay rate and a Higgs
that merely mixes with a weakly coupled singlet. Finding evidence
for a non-SM rate observable using the invisible decay rate of the
Higgs would be beneficial.  However, in many cases
just trying to see evidence for an invisibly
decaying Higgs boson with 100\% branching ratio
is challenging enough~\cite{Frederiksen:me} 
without contemplating measuring a rate observable
associated with a smaller invisible branching ratio 
with sufficient precision to
confirm $R_{ij}(h)=R^{\rm SM}_{ij}(h)(1-B_{\rm inv})$. 

The other way to distinguish between ideas at a hadron collider is
to find the heavier Higgs state.  The SM Higgs rate observables for
this state would be the same as the SM Higgs boson of the
same mass except rescaled by a factor $\sin^2\omega$.  Precisely
measuring this rescaling factor for the heavier Higgs
boson and adding it to the rescaling factor in eq.~\ref{omega rate}
would sum to 1.  The heavier Higgs state would
almost assuredly not be accessible at the Tevatron.  
If it is accessible at the LHC, and depending on its mass, one might be able
to make these measurements with sufficient precision
to test the weak singlet mixing hypothesis.

At an $e^+e^-$ linear collider, there exist
straightforward observables to distinguish $\sin^2\omega$ from
$B_{\rm inv}$ in the rate observables of
eqs.~\ref{inv rate} and~\ref{omega rate}.  A most effective
one is the measurement of the total cross-section for
$e^+e^-\to hZ$.  One identifies the $Z$ from reconstructing
its mass in $Z\to f\bar f$ decays, and measures the recoil spectrum from
the beam constraint~\cite{linear collider}.
Everything from the $h$ decays, whether visible or invisible, will show up
as a peak at $m^2_{\rm recoil}=m_h^2$. This is not a 
$\sigma_i(h)\cdot B_j(h)$ rate
observable, but rather a pure cross-section observable which would
be unaffected by invisible decays of the Higgs but would
be suppressed by singlet mixing.  

The percent-level measurements of
this cross-section and the myriad of other Higgs boson observables 
would easily confirm universal suppression to a compelling level
and also distinguish between the possible reasons behind it.
Furthermore, once the Higgs mass and $\sin^2\omega$ were known to high
precision, one could conceivably combine it with
the precision electroweak data to do an analysis similar to 
what was done to produce Fig.~\ref{s2omega} and identify a new
mass scale at higher energies for future colliders.

\bigskip

\noindent
{\it Acknowledgments: }
JDW was supported in part by the U.S.~Department of
Energy and the Alfred P. Sloan Foundation.

\def\Journal#1#2#3#4{{#1} {\bf #2}, #3 (#4)}
\def\add#1#2#3{{\bf #1}, #2 (#3)}

\def\NPB{{\em Nucl. Phys.} B}
\def\PLB{{\em Phys. Lett.}  B}
\def\PRL{{\em Phys. Rev. Lett.}}
\def\PRD{{\em Phys. Rev.} D}
\def\PR{{\em Phys. Rev.}}
\def\ZPC{{\em Z. Phys.} C}
\def\SJNP{{\em Sov. J. Nucl. Phys.}}
\def\AnnP{{\em Ann. Phys.}}
\def\JETPL{{\em JETP Lett.}}
\def\LMP{{\em Lett. Math. Phys.}}
\def\CMP{{\em Comm. Math. Phys.}}
\def\PTP{{\em Prog. Theor. Phys.}}

\end{document}